\documentclass[sigconf]{acmart}
\AtBeginDocument{%
  }

\setcopyright{acmlicensed}
\copyrightyear{2018}
\acmYear{2018}
\acmDOI{XXXXXXX.XXXXXXX}
\acmConference[Conference acronym 'XX]{Make sure to enter the correct
  conference title from your rights confirmation email}{June 03--05,
  2018}{Woodstock, NY}
\acmISBN{978-1-4503-XXXX-X/2018/06}

\usepackage{microtype}
\usepackage{url}
\usepackage{booktabs}
\usepackage{graphicx}
\usepackage{lineno}
\usepackage{enumitem}
\usepackage{listings} %
\usepackage{svg}
\usepackage{hyperref}

\definecolor{ darkblue}{rgb}{0, 0, 0.5}
\hypersetup{colorlinks=true, citecolor=darkblue, linkcolor=darkblue, urlcolor=darkblue}

\usepackage{multirow}
\usepackage{bigdelim}
\usepackage{todonotes}
\usepackage{longtable}
\usepackage{tabularray}
\usepackage{wrapfig}
\usepackage[most]{tcolorbox} 
\usepackage{url}
\usepackage{xspace}
\usepackage{svg}
\usepackage[absolute]{textpos} %

\usepackage{fdsymbol}   %

\usepackage[utf8]{inputenc} %
\usepackage[T1]{fontenc}    %
\usepackage{url}            %
\usepackage{booktabs}       %
\usepackage{nicefrac}       %
\usepackage{microtype}      %
\usepackage[table]{xcolor}         %
\usepackage[most]{tcolorbox}
\usepackage{csquotes}
\usepackage{wrapfig}
\usepackage{siunitx}
\usepackage{graphicx}
\usepackage{arydshln}
\usepackage{wrapfig}
\usepackage{enumitem}
\usepackage{soul} %

\usepackage{multirow}
\usepackage{xspace}
\usepackage{adjustbox}
\usepackage{pifont}
\usepackage{caption}
\usepackage{makecell}
\usepackage{subcaption}
\usepackage{bold-extra}
\usepackage{url}
\usepackage{float}
\usepackage{pgf-pie}
\usepackage{balance}

\usepackage[normalem]{ulem}

\usepackage{cleveref}
\crefname{figure}{Figure}{Figures}
\crefname{section}{Section}{Sections}
\crefname{equation}{Equation}{Equations}
\crefname{appendix}{Appendix}{Appendice}
\crefname{table}{Table}{Tables}

\definecolor{linkcolor}{RGB}{0, 0, 128}
\hypersetup{
     colorlinks   = true,
     citecolor    = linkcolor,
     linkcolor    = linkcolor,
     urlcolor     = linkcolor,
}
\usepackage{pifont}%
\usepackage{listings}

\usepackage{setspace}

\usepackage{nicematrix}
\newcolumntype{L}[1]{>{\raggedright\let\newline\\\arraybackslash\hspace{0pt}}m{#1}}
\newcolumntype{C}[1]{>{\centering\let\newline\\\arraybackslash\hspace{0pt}}m{#1}}
\newcolumntype{R}[1]{>{\raggedleft\let\newline\\\arraybackslash\hspace{0pt}}m{#1}}
\newcolumntype{P}[1]{>{\centering\let\newline\\\arraybackslash\hspace{0pt}}m{#1}}
\newcommand{\sans}[1]{\textsf{#1}}

\newcommand{\ours}{\texttt{LACONIC}\xspace}
\newcommand{\oursOneB}{\texttt{LACONIC-1B}\xspace}
\newcommand{\oursThreeB}{\texttt{LACONIC-3B}\xspace}
\newcommand{\oursEightB}{\texttt{LACONIC-8B}\xspace}

\newcommand{\splade}{\texttt{SPLADE}\xspace}
\newcommand{\csplade}{\texttt{CSPLADE}\xspace}
\newcommand{\llama}{\texttt{Llama3}\xspace}
\newcommand{\nomic}{\texttt{Nomic-text-v1}\xspace}
\newcommand{\arctic}{\texttt{Arctic-embed-v2}\xspace}
\newcommand{\seismic}{\textsc{Seismic}\xspace}

\newcommand{\huggingface}{\raisebox{-1.5pt}{\includegraphics[height=1.05em]{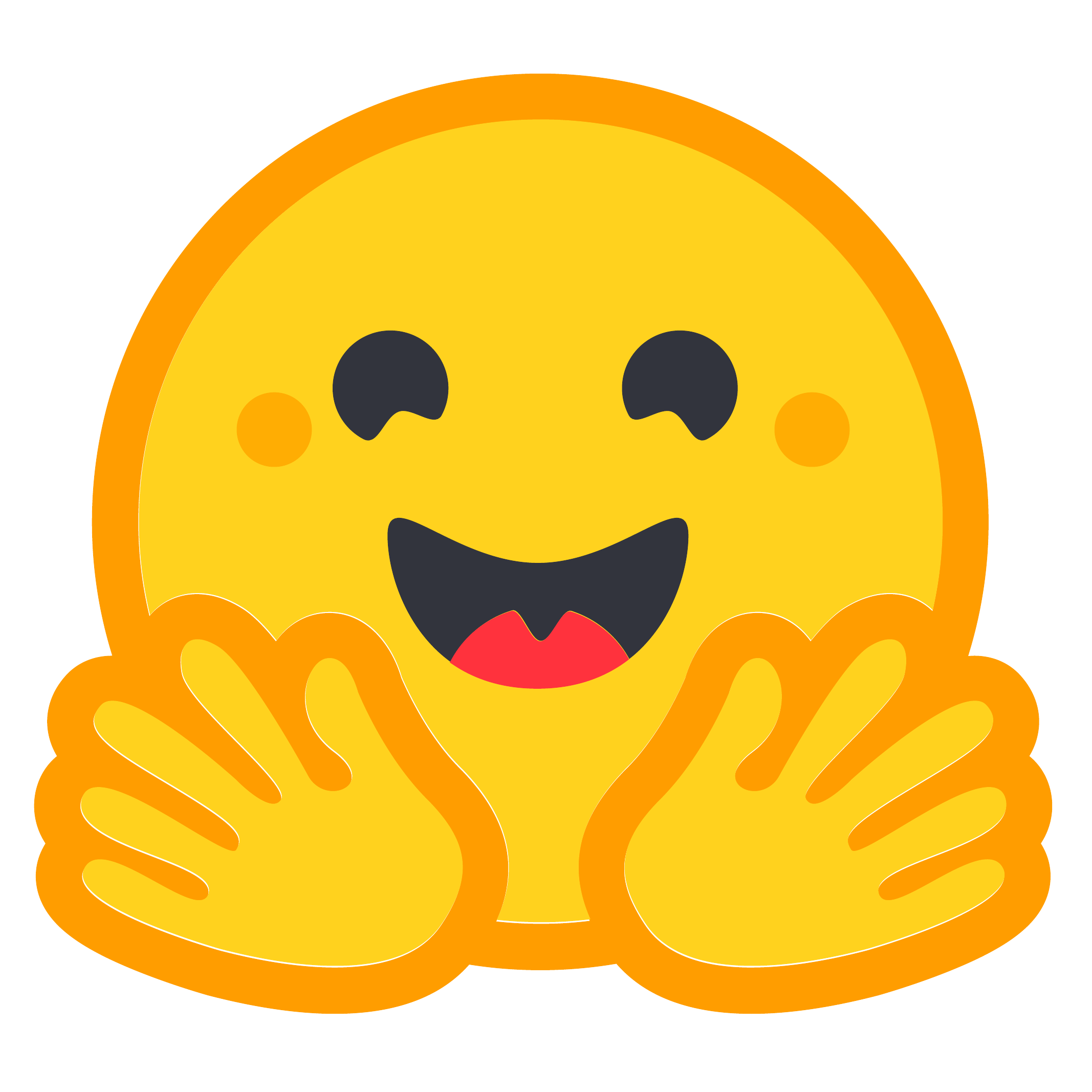}}\xspace}

\newcommand{\github}{\raisebox{-1.5pt}{\includegraphics[height=1.05em]{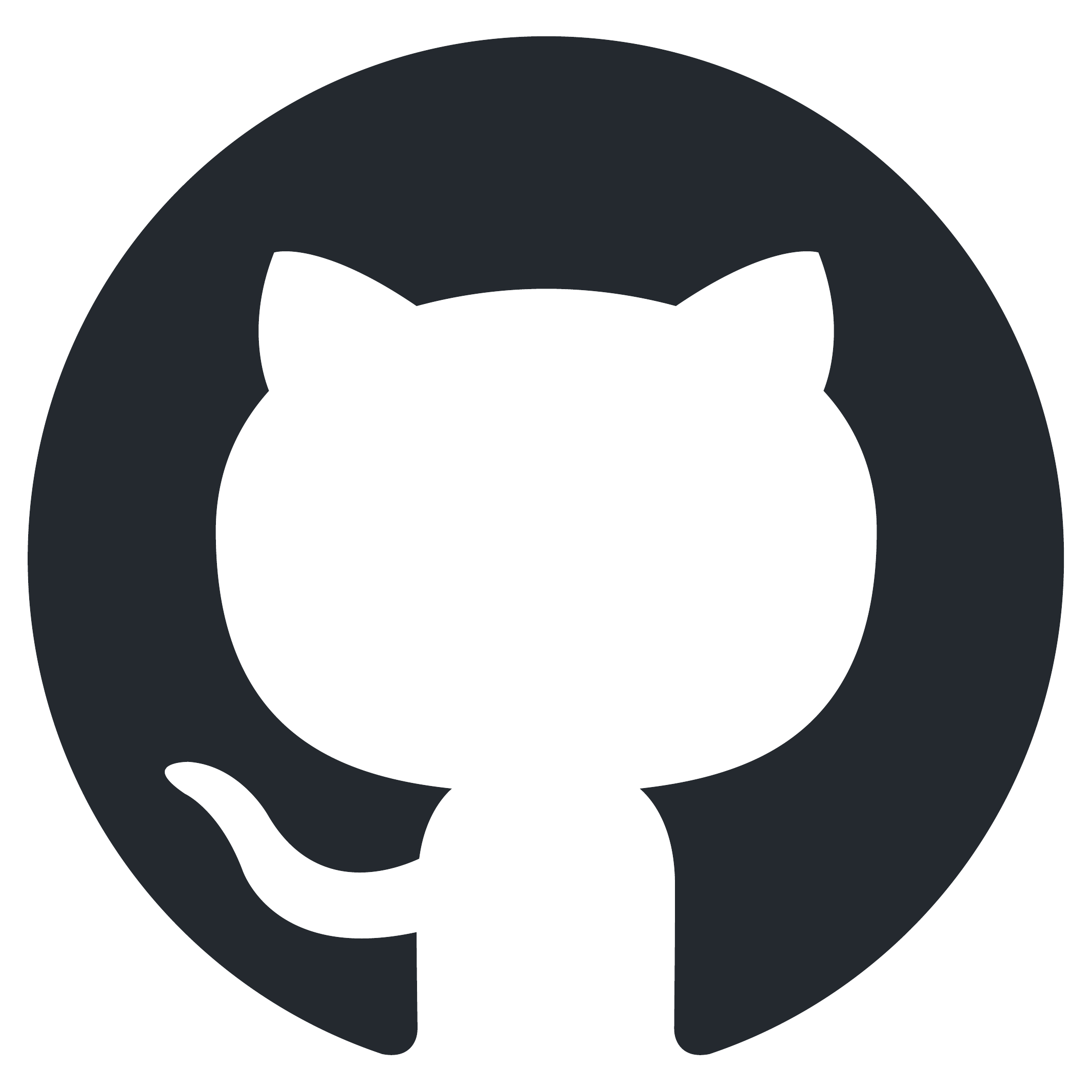}}\xspace}

\copyrightyear{2026}
\acmYear{2026}
\setcopyright{cc}
\setcctype{by}
\acmConference[SIGIR '26]{Proceedings of the 49th International ACM SIGIR Conference on Research and Development in Information Retrieval}{July 20--24, 2026}{Melbourne, VIC, Australia}
\acmBooktitle{Proceedings of the 49th International ACM SIGIR Conference on Research and Development in Information Retrieval (SIGIR '26), July 20--24, 2026, Melbourne, VIC, Australia}
\acmDOI{10.1145/3805712.3809869}
\acmISBN{979-8-4007-2599-9/2026/07}

\begin{document}

\title{LACONIC: Dense-Level Effectiveness for Scalable Sparse Retrieval via a Two-Phase Training Curriculum}

% \affiliation{%

\author{Zhichao Xu}
\email{zhichao.xu@utah.edu}
\affiliation{%
  \institution{University of Utah}
  \city{Salt Lake City}
  \state{UT}
  \country{USA}
}

\author{Shengyao Zhuang}
\email{s.zhuang@uq.edu.au}
\affiliation{%
  \institution{The University of Queensland}
  \city{Brisbane}
  \state{QLD}
  \country{Australia}
}

\author{Crystina Zhang}
\email{x978zhan@uwaterloo.ca}
\affiliation{%
  \institution{University of Waterloo}
  \city{Waterloo}
  \state{ON}
  \country{Canada}
}

\author{Xueguang Ma}
\email{x93ma@uwaterloo.ca}
\affiliation{%
  \institution{University of Waterloo}
  \city{Waterloo}
  \state{ON}
  \country{Canada}
}

\author{Yijun Tian}
\email{meetyijun@gmail.com}
\affiliation{%
  \institution{University of Notre Dame}
  \city{Notre Dame}
  \state{IN}
  \country{USA}
}

\author{Maitrey Mehta}
\email{maitrey@cs.utah.edu}
\affiliation{%
  \institution{University of Utah}
  \city{Salt Lake City}
  \state{UT}
  \country{USA}
}

\author{Jimmy Lin}
\email{jimmylin@uwaterloo.ca}
\affiliation{%
  \institution{University of Waterloo}
  \city{Waterloo}
  \state{ON}
  \country{Canada}
}

\author{Vivek Srikumar}
\email{svivek@cs.utah.edu}
\affiliation{%
  \institution{University of Utah}
  \city{Salt Lake City}
  \state{UT}
  \country{USA}
}

\renewcommand{\shortauthors}{Zhichao Xu et al.}

\begin{abstract}
  While dense retrieval models have been the standard for state-of-the-art information retrieval, their deployment is often constrained by high memory requirements and reliance on GPU accelerators for vector similarity search at scale. Learned sparse retrieval offers a compelling alternative by enabling efficient search via inverted indices, yet it has historically received less attention than dense approaches. In this paper, we introduce \ours, a family of learned sparse retrievers based on the \llama~architecture (1B, 3B, and 8B). We propose a streamlined two-phase training curriculum consisting of (1) weakly supervised pre-finetuning to adapt causal LLMs for bidirectional contextualization and (2) high-signal finetuning using curated hard negatives. Our results demonstrate that \ours~effectively bridges the performance gap with dense models: the 8B variant achieves a state-of-the-art 60.2 nDCG@10 on the MTEB Retrieval benchmark, ranking 15th on the leaderboard as of February 5th, 2026, while utilizing 74\% less index memory than an equivalent dense model. By delivering high retrieval effectiveness on commodity CPU hardware with a fraction of the compute budget required by competing models, \ours~provides a scalable and efficient solution for real-world search applications. We fully open source our code implementation and trained checkpoints to facilitate reproducibility.
\end{abstract}

\begin{CCSXML}
<ccs2012>
   <concept>
       <concept_id>10002951.10003317.10003338</concept_id>
       <concept_desc>Information systems~Retrieval models and ranking</concept_desc>
       <concept_significance>500</concept_significance>
       </concept>
 </ccs2012>
\end{CCSXML}

\ccsdesc[500]{Information systems~Retrieval models and ranking}

\keywords{Learned sparse retrieval; document representation.}

\maketitle

\begin{center}
\small

\begin{minipage}{0.95\columnwidth}
\centering

\github~{\sans{Code}}~
\href{https://github.com/zhichaoxu-shufe/laconic-sparse-retrieval}{\texttt{laconic-sparse-retrieval}} \\[3pt]

\huggingface~{\sans{Data}}~
\href{https://huggingface.co/datasets/utahnlp/nomic-embed-pretrain-lite}{\texttt{nomic-embed-pretrain-lite}} \\[3pt]

\huggingface~{\sans{Models}}~
\href{https://huggingface.co/utahnlp/laconic-1b}{\oursOneB} \quad
\href{https://huggingface.co/utahnlp/laconic-3b}{\oursThreeB} \quad
\href{https://huggingface.co/utahnlp/laconic-8b}{\oursEightB}

\end{minipage}

\end{center}

\section{Introduction}
Information retrieval (IR) has undergone a paradigm shift from traditional term-matching methods like BM25~\citep{robertson1995okapi} to neural dense retrieval models~\citep{lee-etal-2019-latent,karpukhin-etal-2020-dense,lin2022pretrained,xu2025surveymodelarchitecturesinformation}. Dense retrievers excel at capturing semantic nuances by encoding queries and documents into continuous high-dimensional vectors. However, they often suffer from significant deployment overhead, requiring large memory footprints to store dense embeddings and specialized hardware (e.g., GPUs) for efficient vector similarity search.

Learned sparse retrieval, pioneered by earlier works such as \texttt{SNRM}~\citep{zamani2018neural}, \texttt{DeepCT}~\citep{dai2019deeper}, \texttt{SparTerm}~\citep{bai2020sparterm}, \texttt{SPARTA}~\citep{zhao-etal-2021-sparta} and popularized by the \splade framework~\citep{formal2021spladev1,formal2021spladev2,formal2022spladeplus,formal2022efficientsplade,lassance2024spladev3,joelsplade}, offers a compelling middle ground. By projecting hidden states onto the vocabulary space and applying sparsity-inducing regularizations, these models produce high-dimensional but sparse representations. This allows for the use of efficient, CPU-friendly inverted index structures while maintaining the semantic richness of neural encoders. Despite their potential, a performance gap has historically persisted between sparse models and state-of-the-art dense retrievers, particularly as the latter have scaled to large language model (LLM) backbones~\citep{zhu2023large}.

In this paper, we introduce \ours, a series of learned sparse retrieval models based on the \llama family~\citep{grattafiori2024llama3technicalreport} in 1B, 3B, and 8B parameter scales. We name our model \ours as a tribute to the historical tradition of \textit{laconic} speech\,---\,the Spartan practice of using the fewest words possible to deliver the maximum impact. This serves as a technical metaphor for our architecture: we leverage the vast knowledge of autoregressive decoders but constrain them to generate succinct, vocabulary-sparse representations that are ``Spartan'' in their resource requirements.

We adopt a streamlined two-phase training curriculum\,---\,pre-finetuning on weakly-supervised data followed by high-quality hard-negative finetuning\,---\,that allows \ours to close the performance gap with its dense counterparts. 
Although such multi-stage curricula have been extensively studied and validated in dense retrieval, we show that this paradigm is equally critical for learned sparse retrievers, where it plays a central role in adapting large causal language models to bidirectional information and relevance modeling.
In contrast to \splade-v2's curriculum~\citep{formal2021spladev2}, which couples hard-negative mining with cross-encoder distillation atop a BERT backbone, our recipe targets the distinct challenges of adapting large causal LLMs for bidirectional sparse encoding without auxiliary teacher models, complementing recent decoder-based LSR efforts~\citep{doshi2024mistralspladellmsbetterlearned,qiao2025leveragingdecoderarchitectures,zeng2025scalingsparseanddense,xu2025cspladelearnedsparseretrieval,formal2026learningretrievalmodelssparse}.
Our \ours-8B model achieves an impressive 60.2 nDCG@10 on the MTEB Retrieval benchmark, ranking 15th on the leaderboard as of February 5th, 2026. Notably, \ours achieves these results using a fraction of the compute budget of its competitors while maintaining a significantly smaller index memory footprint. We detail our architecture, training methodology, an extensive evaluation of retrieval efficiency and effectiveness, and fully open source our implementation and trained models.

\section{Proposed Approach}
The performance of \ours stems from the integration of powerful LLM backbones with a streamlined two-phase training curriculum based on contrastive training.

\subsection{Model Architecture}
\label{subsec:model_architecture}
\ours is a bi-encoder retrieval model~\citep{reimers-gurevych-2019-sentence,Humeau2020Poly-encoders}, which builds upon the \splade framework~\citep{formal2021spladev1,formal2022efficientsplade} while incorporating architectural best practices for scaling sparse retrievers~\citep{doshi2024mistralspladellmsbetterlearned,qiao2025leveragingdecoderarchitectures,xu2025cspladelearnedsparseretrieval}.

Denote query $Q$ and document $D$, and a language model's vocabulary as $\mathcal{V}$.
$D=\{t_1, t_2, \ldots, t_{|D|}\}$ where $t_i$ is the $i$-th token. The document's corresponding contextualized representation can be written as $\{\boldsymbol{h}_1, \boldsymbol{h}_2, \ldots \boldsymbol{h}_{|D|} \}$. For each $\boldsymbol{h}_i$, we project the hidden representation to a vocabulary-sized vector $\boldsymbol{H}_i \in \mathbb{R}^{|\mathcal{V}|}$ with the language modeling head.
The $j$-th dimension of $\boldsymbol{H}_i$ represents the importance of token $j$ (in vocabulary $\mathcal{V}$) to token $i$ in the input sequence, which in practice is the $\operatorname{logit}_j$ from the LM head output.
Given $\boldsymbol{H}_{D} = \{\boldsymbol{H}_1, \boldsymbol{H}_2, \ldots \boldsymbol{H}_{|D|} \}$ of tensor shape $(|\mathcal{V}|, |D|)$, we apply a max-pooling along the sequence length dimension, i.e., across all tokens, followed by ReLU activation and log rescaling to get the vocabulary-sized representation for the input document $d$:
\begin{equation}
    \boldsymbol{D} = \log \Big( 1+\operatorname{ReLU} \big( \operatorname{MaxPooling} (\boldsymbol{H}_D) \big) \Big) \in \mathbb{R}^{|\mathcal{V}|}
    \label{eq:transformation}
\end{equation}
A similar operation can also be applied to query $Q$ to get query representation $\boldsymbol{Q}\in \mathbb{R}^{|\mathcal{V}|}$.

We adopt~\llama family models as the backbone, specifically \texttt{Llama3ForCausalLM}. Note that the above~\cref{eq:transformation} applies a pooling along the sequence length dimension, which is disadvantageous for causal language models with unidirectional attention. Prior works explored different mitigations, including using ``echo'' input~\citep{springer2024echoembedding,doshi2024mistralspladellmsbetterlearned} and enabling bidirectional attention via lightweight adaptation training~\citep{behnamghader2024llmvec,zeng2025scalingsparseanddense,xu2025cspladelearnedsparseretrieval}; in contrast, we adopt a streamlined approach of directly enabling the bidirectional attention of the causal language models by removing the causal attention mask, and letting the models ``self-adapt'' in the subsequent contrastive training.

To summarize, \ours differs from~\splade by using a bidirectional variant of the stronger~\llama backbone language model, which is implementation-wise straightforward and achieves impressive empirical performance with a correct training curriculum.

\subsection{Training Objective}
We adopt a standard InfoNCE loss~\citep{oord2018representation} in training~\ours. Denote a training pair $(Q, D^{+})$, where $D^{+}$ is relevant to query $Q$, and $\{D_N \}$ is a list of documents not relevant to $Q$, score function $s(Q,D)=\langle\boldsymbol{Q}, \boldsymbol{D}\rangle$, the ranking loss is formulated as: 
\begin{align}
    \mathcal{L}_{rank}(Q, D^{+}, \{D_N\})
    &= -\log p(D=D^+ | Q) \notag \\
    &= - \log \frac{e^{s(Q, D^+)}}{e^{s(Q, D^+)} + \sum\limits_{D_i^- \in \{D_N \}} e^{s(Q, D_i^-)}}
    \label{eq:infonce}
\end{align}
Practically we use in-batch negatives and/or hard negatives in different phases of training, following prior practices~\citep{nussbaum2025nomicembedtrainingreproducible,yu2024arcticembed20multilingualretrieval}.
To enforce the sparsity of the encoded sparse representations, we adopt \texttt{FLOPs} regularization~\citep{paria2020minimizingflops}, the same as \splade. 
Denote \texttt{FLOPs} regularization loss for $Q$ and $D$ as $\mathcal{L}_{reg}^Q$ and $\mathcal{L}_{reg}^D$, respectively, and $\lambda_Q$, $\lambda_D$ as the corresponding coefficients, the final loss is:
$$
\mathcal{L} = \mathcal{L}_{rank}(Q, D^{+}, \{D_N\}) + \lambda_Q \mathcal{L}_{reg}^Q + \lambda_D \mathcal{L}_{reg}^D 
$$
where $\lambda_Q$ and $\lambda_D$ are tuned as hyperparameters.

\subsection{Pre-finetuning}
The goal of this training phase is to adapt the backbone language model for bidirectional attention (\cref{subsec:model_architecture}) while training it to encode sparse representations to model query and document relevance using large-scale, weakly-supervised data. 

\paragraph{Dataset}
We follow prior recipes~\citep{günther2024jinaembeddings28192token,nussbaum2025nomicembedtrainingreproducible,yu2024arcticembed20multilingualretrieval} to use weakly-supervised contrastive pairs, i.e., $(Q,D)$ pairs curated from noisy data sources. Given the limited compute budget, we use a subset of Nomic Embedding Unsupervised Data released under Apache 2.0 license.\footnote{\url{https://huggingface.co/datasets/nomic-ai/nomic-embed-unsupervised-data}}
As our focus is on asymmetric retrieval tasks, we use 11 splits: \texttt{wikipedia}, \texttt{gooaq}, \texttt{agnews}, \texttt{ccnews}, \texttt{npr}, \texttt{eli5}, \texttt{cnn}, \texttt{squad}, \texttt{quora}, \texttt{simplewiki}, \texttt{stackexchange\_duplicate\_questions}. We selected these splits via lightweight manual inspection, informed by our compute budget. Our final mixture consists of about 9M pairs, which is a small fraction of the original dataset's 470M pairs or \texttt{Arctic-Embed-v2}'s 308M pairs~\citep{yu2024arcticembed20multilingualretrieval}. We refer to this mixture as \texttt{Nomic-embed-pretrain-lite}. We hypothesize that scaling the pre-finetuning data could further improve performance.

\paragraph{Training}
In this training phase, we use in-batch negatives. Prior works have reported the efficacy of scaling up batch size in contrastive training~\citep{gao-etal-2021-scaling}. Notably, \nomic~\citep{nussbaum2025nomicembedtrainingreproducible} used 16,384 global batch size while \arctic~\citep{yu2024arcticembed20multilingualretrieval} reported 32,768 global batch size using 32 $\times$ H100 GPUs. In our experiments, we use 2,048 global batch size consistently per our compute budget. 

We carefully tune the training schedule and hyperparameters. We train for 3, 2, 1 epochs for 1B, 3B, and 8B variants of \ours, respectively. We use LoRA training~\citep{hu2021lora} and set rank=32 for 1B and 3B models, and rank=16 for the 8B model. We use cosine learning rate scheduling, and adopt a separate exponential warmup for the \texttt{FLOPs} regularization loss, same as~\citep{formal2021spladev1,formal2021spladev2}. We set $\lambda_{Q}=\lambda_{D}=1\times 10^{-3}$ for \{1B, 3B, 8B\} models. We truncate the queries to 64 tokens and documents to 192 tokens. 
After this training phase, we merge the LoRA adapter back to the base model and use the merged checkpoint in the subsequent finetuning phase. 

\subsection{Finetuning}
After the pre-finetuning phase, the model has already learned the sparsity pattern required for sparse retrieval and acquired the basic ``capability'' of identifying relevance. We then move on to the next phase of finetuning with dedicated hard negatives.

\begin{table*}[t]
\centering
\caption{Model Performance (nDCG@10). We use baseline results reported by their respective papers. Average BEIR14* excludes the result on CQADupstack datasets.}
\label{tab:results}
\small
\setlength{\tabcolsep}{3pt} % 进一步压缩列间距
\begin{tabular}{llc ccccccccccccccc cc}
\toprule
\textbf{Category} & \textbf{Model} & \textbf{Size} & \textbf{MS} & \textbf{Arg} & \textbf{Clm} & \textbf{CQA} & \textbf{DBP} & \textbf{FEV} & \textbf{FiQ} & \textbf{Hot} & \textbf{NFC} & \textbf{NQ} & \textbf{Quo} & \textbf{SCI} & \textbf{ScF} & \textbf{TRC} & \textbf{Tou} & \textbf{Avg BEIR14} & \textbf{MTEB} \\
\midrule
\multirow{2}{*}{\shortstack[l]{Sparse}} 
 & SPLADE-v3 & 110M & 45.6 & 50.9 & 23.3 & 32.3 & 45.0 & 79.6 & 37.4 & 69.2 & 35.7 & 58.6 & 81.4 & 15.8 & 71.0 & 74.8 & 29.3 & 51.3 & 50.0 \\
 & CSPLADE & 8B & 46.5 & 48.9 & 29.4 & -- & 44.5 & 86.5 & 40.5 & 69.8 & 37.2 & 60.9 & 87.1 & 17.6 & 73.9 & 83.2 & 38.9 & 54.6 & -- \\
\midrule
\multirow{3}{*}{\shortstack[l]{Dense}} 
 & Nomic-v1 & 137M & 43.1 & 49.3 & 40.5 & 38.3 & 45.0 & 85.0 & 38.4 & 73.6 & 35.0 & 59.4 & 87.7 & 18.3 & 70.5 & 79.9 & 28.2 & 53.9 & 52.8 \\
 & Arctic-v2 & 529M & 44.0 & 58.0 & 38.3 & 47.2 & 43.9 & 91.6 & 44.0 & 72.4 & 35.9 & 64.6 & 88.7 & 20.3 & 71.8 & 80.3 & 29.8 & 56.0 & 55.4 \\
 & RepLlama3 & 8B & 45.3 & 60.2 & 42.3 & 44.2 & 45.8 & 91.8 & 57.3 & 85.0 & 41.7 & 70.1 & 85.9 & 29.6 & 78.6 & 85.8 & 33.7 & \underline{60.9} & \underline{59.8} \\
\midrule
\multirow{3}{*}{\textbf{Ours}} 
 & \oursOneB & 1B & 43.5 & 62.1 & 37.4 & 34.3 & 46.8 & 88.3 & 43.4 & 79.2 & 39.2 & 66.3 & 85.2 & 24.2 & 75.6 & 83.9 & 31.2 & 57.6 & 56.0 \\
 & \oursThreeB & 3B & 44.0 & 72.0 & 36.4 & 41.5 & 49.0 & 88.5 & 50.7 & 81.7 & 41.0 & 69.8 & 86.7 & 27.3 & 78.2 & 83.4 & 30.7 & 60.0 & 58.7 \\
 & \oursEightB & 8B & 44.1 & 73.0 & 38.8 & 42.3 & 50.2 & 89.8 & 55.0 & 83.9 & 41.9 & 72.8 & 86.1 & 29.3 & 79.7 & 85.3 & 31.3 & \textbf{61.5} & \textbf{60.2} \\
\bottomrule
\end{tabular}
\end{table*}
\begin{figure*}[t!]
\centering
\includegraphics[width=1\textwidth]{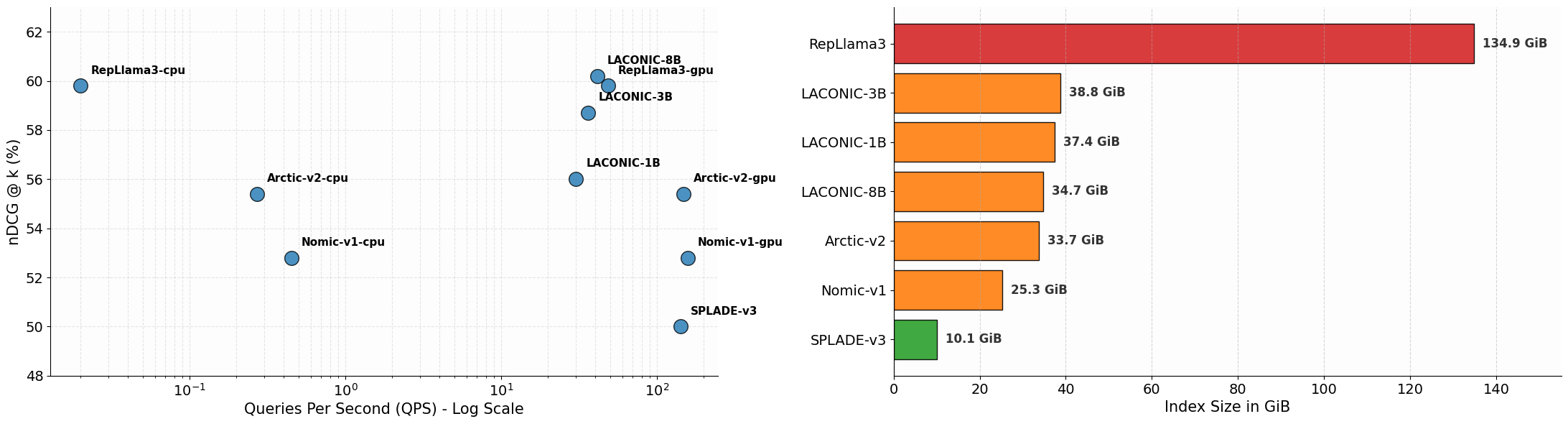}
\caption{Efficiency comparison. Left plot shows the index search latency on MS MARCO dataset, measured by queries per second, versus retrieval performance on MTEB-R benchmark. Right plot shows memory requirement to load retrieval index. Notice that \ours improves the performance-latency frontier compared to baselines without requiring accelerators for efficient index search. We reproduce \texttt{SPLADE-v3}'s result using \seismic.}
\label{fig:index-efficiency}
% \vspace{0pt}
\end{figure*}

\paragraph{Dataset}
We adopt a recently released \texttt{RLHN} dataset~\citep{thakur-etal-2025-hard}, which consists of 690K $(Q, D^+, \{D_N\})$ triplets.\footnote{\url{https://huggingface.co/datasets/rlhn/rlhn-680K}} \texttt{RLHN} is a lite version of the larger BGE training mixture~\citep{li2025makingtextembedders} with further relabeled hard negatives, and has been reported to improve retrieval and ranking performance while reducing training time~\citep{thakur-etal-2025-hard,xu2025distillationversuscontrastivelearning}. More specifically, the BGE mixture's negatives are mined as top-ranked but non-relevant candidates by an ensemble of strong dense and sparse retrievers; \texttt{RLHN} then re-judges these candidates with an LLM judge to remove false negatives, yielding cleaner per-query hard negatives than typical mining-only pipelines.

\paragraph{Training}
We use hard negatives and in-batch negatives in this training phase. Specifically, each query is paired with 1 relevant document and 15 hard negatives, together with all other documents in this batch. We use a consistent 32 global batch size, which implies 512 negatives per query. Similar to the pre-finetuning phase, we finetune for 2, 2, 1 epochs for 1B, 3B, and 8B variants of \ours. Again, we use LoRA training, set rank=32 for 1B and 3B models and rank=16 for the 8B model, together with cosine learning rate scheduling, a separate exponential warmup for \texttt{FLOPs} regularization loss and use a consistent $\lambda_{Q}=\lambda_{D}=1 \times {10}^{-3}$. We truncate both queries and documents to 192 tokens.

\section{Experiments}
We describe the experimental setup (\cref{subsec:experiment_setup}) and discuss results and analysis (\cref{subsec:results}).
\subsection{Experimental Setup}
\label{subsec:experiment_setup}
\paragraph{Implementation Details}
We initialize \ours with open-weight \texttt{Llama3.2-1B},\footnote{\url{https://huggingface.co/meta-llama/Llama-3.2-1B}} \texttt{Llama3.2-3B},\footnote{\url{https://huggingface.co/meta-llama/Llama-3.2-3B}} \texttt{Llama3.1-8B}\footnote{\url{https://huggingface.co/meta-llama/Llama-3.1-8B}} models, all licensed for academic use. Note that we use the Base models (without post-training).
We implement \ours in PyTorch on top of the \texttt{Tevatron} framework~\citep{gao2022tevatron}. To improve training scalability, we use gradient checkpointing, gradient accumulation, mixed-precision training (BF16), Flash Attention 2~\citep{dao2023flashattention}, and PyTorch FSDP~\citep{zhao2023pytorchdsdp}.
Our compute infrastructure is based on a cluster of A100 SXM4 40GB GPUs with NVSwitch inter-GPU connectivity.

\paragraph{Baselines}
For dense retrieval models, we include open-weight models: \texttt{Nomic-Embed-v1}~\citep{nussbaum2025nomicembedtrainingreproducible}\footnote{\url{https://huggingface.co/nomic-ai/nomic-embed-text-v1}}, \texttt{Arctic-Embed-v2}~\citep{yu2024arcticembed20multilingualretrieval}.\footnote{\url{https://huggingface.co/Snowflake/snowflake-arctic-embed-l-v2.0}}
We also implement a dense baseline following \texttt{RepLlama}'s training recipe with the \texttt{Llama3.1-8B} backbone and \texttt{RLHN} dataset, which we term \texttt{RepLlama3}. Note that \texttt{RepLlama3} did not undergo the same pre-finetuning training phase as~\ours. 
For sparse retrieval models, we include the performant \texttt{SPLADE-v3}~\citep{lassance2024spladev3} and \texttt{CSPLADE}~\citep{xu2025cspladelearnedsparseretrieval}.

\paragraph{Indexing and Evaluation} We use \seismic\,--\,an efficient inverted index structure~\citep{bruch2024efficient,bruch2025efficientsketchingnearestneighbor} that enables accurate and fast approximate nearest neighbor search without GPU accelerators. We also compare retrieval latency against dense baselines using a Faiss FlatIP dense index~\citep{johnson2017billionscalesimilaritysearchgpus}.

We use nDCG@10 to evaluate \ours's retrieval performance on the 15 retrieval tasks of the MTEB benchmark~\citep{muennighoff-etal-2023-mteb}, which we refer to as MTEB-R.

\subsection{Results and Analysis}
\label{subsec:results}

\paragraph{Retrieval Performance}
We report the retrieval performance in~\cref{tab:results}. The lite \oursOneB significantly outperforms the state-of-the-art sparse retrieval models \splade and \csplade, averaging 57.6 nDCG@10 on 14 BEIR datasets versus \splade's 51.3 and \csplade's 54.6. \oursOneB also outperforms the competitive lightweight \nomic and \arctic, despite being trained on only a fraction of pre-finetuning and finetuning datasets. Our largest model, \oursEightB achieves an impressive average 60.2 nDCG@10 on 15 MTEB Retrieval datasets, which is ranked 15th on the leaderboard as of February 5th, 2026, being the only learned sparse retrieval model at this position. We note that on the MS MARCO development set specifically, \ours trails \splade-v3 and \csplade by approximately 1.5--3.0 nDCG@10. We attribute this to \splade-v3 and \csplade being trained extensively on MS MARCO supervision with cross-encoder distillation, whereas \ours's training mixture targets broad-domain generalization without MS MARCO-specific tuning. The substantial gains on the cross-domain MTEB Retrieval datasets suggest this trade-off favors generalization, while domain-specific finetuning could further improve \ours's performance on individual benchmarks.

\paragraph{Index Search Efficiency}

We study the index search efficiency of \ours compared to dense and sparse baselines, with results shown in \cref{fig:index-efficiency}. Unless otherwise stated, we report query-time search latency only, excluding embedding computation and index construction costs. For dense retrieval, we use Faiss GPU index (\texttt{GpuIndexFlatIP}) with 8× A100 SXM4 40GB GPUs or CPU-only \texttt{IndexFlatIP}, while \ours uses the \seismic inverted index on CPU without accelerator support. All benchmarks are conducted on a compute node with an Intel Xeon Platinum 8275L CPU and 1152 GB RAM.

\begin{figure}[t]
\centering
\includegraphics[width=1\columnwidth]{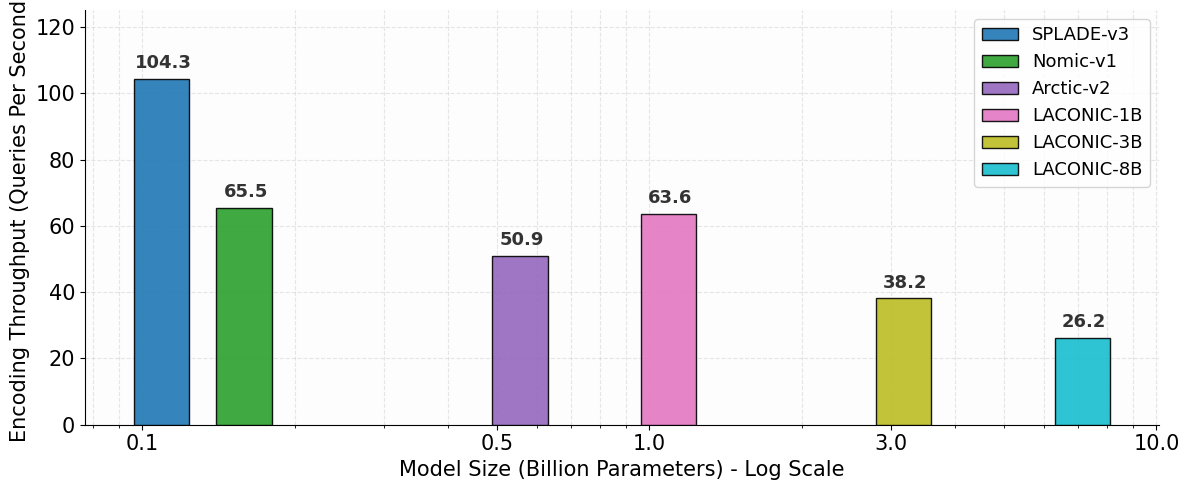}
\caption{Encoding speed comparison at batch\_size=1.}
\label{fig:encoding-speed}
\vspace{0pt}
\end{figure}

As shown in the left panel, \ours achieves a superior effectiveness–latency trade-off compared to dense baselines, enabling fast approximate nearest neighbor search using inverted indices alone. Notably, \ours does not require GPU accelerators at index search time, significantly reducing deployment complexity. We note that this latency excludes query encoding, which for an 8B decoder is non-trivial; encoding speed is reported separately in~\cref{fig:encoding-speed}, and we defer a fully integrated end-to-end latency benchmark to future work.

\begin{table*}[t!]
\centering
\caption{Ablation study of dense versus learned sparse retrievers at 1B scale. }
\label{tab:ablation_dense_versus_sparse}
\small
\begin{tabular}{ll ccccccccccccccc c}
\toprule
\textbf{Category} & \textbf{Method} & \textbf{MS} & \textbf{Arg} & \textbf{Clm} & \textbf{CQA} & \textbf{DBP} & \textbf{FEV} & \textbf{FiQ} & \textbf{Hot} & \textbf{NFC} & \textbf{NQ} & \textbf{Quo} & \textbf{SCI} & \textbf{ScF} & \textbf{TRC} & \textbf{Tou} & \textbf{Avg} \\
\midrule
\multirow{2}{*}{Unsupervised} & Contriever & 20.6 & 37.9 & 15.5 & 28.4 & 29.2 & 68.2 & 24.5 & 48.1 & 31.7 & 25.4 & 83.5 & 14.9 & 64.9 & 27.4 & 19.3 & 36.0 \\
 & E5-base & 26.0 & 42.2 & 15.4 & 35.4 & 35.4 & 63.4 & 40.0 & 52.4 & 35.8 & 39.0 & 85.7 & 21.1 & 73.7 & 61.0 & 16.9 & 42.9 \\
\midrule
\multirow{2}{*}{Supervised} & Contriever & 40.7 & 44.6 & 23.7 & 34.5 & 41.3 & 75.8 & 32.9 & 63.8 & 32.8 & 49.8 & 86.5 & 16.5 & 67.7 & 59.6 & 23.0 & 46.2 \\
 & E5-base & 43.1 & 51.4 & 15.4 & 38.9 & 41.0 & 58.2 & 36.4 & 62.2 & 36.6 & 60.0 & 87.9 & 19.0 & 73.1 & 79.6 & 28.3 & 48.7 \\
\midrule
\multirow{2}{*}{Dense} & Pre-FT & 32.4 & 54.8 & 22.9 & 42.0 & 37.6 & 74.7 & 42.9 & 63.1 & 36.6 & 45.5 & 88.3 & 21.2 & 73.7 & 71.5 & 24.5 & 48.8 \\
 & FT & 43.6 & 57.6 & 38.0 & 42.4 & 46.4 & 89.7 & 46.6 & 78.9 & 37.9 & 65.3 & 86.9 & 25.4 & 76.6 & 84.7 & 32.1 & \textbf{56.8} \\
\midrule
\multirow{2}{*}{Sparse} & Pre-FT & 26.6 & 52.5 & 22.1 & 33.4 & 33.1 & 59.9 & 34.0 & 54.9 & 35.7 & 33.6 & 84.5 & 18.4 & 69.5 & 67.0 & 13.3 & 42.6 \\
 & FT & 43.5 & 62.1 & 37.4 & 34.3 & 46.8 & 88.3 & 43.4 & 79.2 & 39.2 & 66.3 & 85.2 & 24.2 & 75.6 & 83.9 & 31.2 & \underline{56.0} \\
\bottomrule
\end{tabular}
\end{table*}

The right panel highlights the memory efficiency of learned sparse indexing. Compared to the dense \texttt{RepLlama3} model, which requires 134.9 GiB to index the corpus, \oursEightB requires only 34.7 GiB\,---\,corresponding to a 3.9$\times$ reduction in index size. This substantially smaller memory footprint enables retrieval on commodity hardware and underscores the practical advantages of learned sparse representations. For a closer comparison within the LSR family, \splade-v3 and \csplade index the same corpus at approximately 10~GiB and >30~GiB, respectively. The roughly $3\times$ overhead of \ours over BERT-scale \splade-v3 is broadly aligned with other LLM-backed LSR systems and reflects the higher token-activation density of LLM encoders rather than an artifact of our training objective; tighter \texttt{FLOPs} regularization scheduling could narrow this gap. We also note that dense indices admit complementary compression techniques such as product quantization (e.g., OPQ) and Matryoshka representation learning (MRL); these are orthogonal to LSR, and a head-to-head comparison at matched memory budgets remains valuable future work.

\paragraph{Encoding Speed}

We report encoding speed (queries per second, QPS) in~\cref{fig:encoding-speed} at BF16 precision. All results are based on a single A100 SXM4 GPU. Compared to baselines, \ours demonstrates competitive QPS due to native support of Flash Attention.

\paragraph{Ablation Studies}

In~\cref{tab:ablation_dense_versus_sparse}, we compare the sparse retrieval model with the dense model that undergoes the same training curriculum. This ablation is carried out at 1B model scale. We include additional baselines \texttt{Contriever}~\citep{izacard2022unsupervised} and \texttt{e5-base}~\citep{wang2022textembeddingbyweaklysupervisedpretraining} (both pre-finetuned and finetuned variants) to showcase the effectiveness of pre-finetuning.

We note the importance of the pre-finetuning phase and using a stronger backbone model. The dense 1B retriever already achieves 48.8 nDCG@10 on MTEB Retrieval datasets, outperforming the finetuned \texttt{e5-base} baseline. We observe that \ours underperforms its dense counterpart after pre-finetuning, which we hypothesize is because the dense model by default uses \#\texttt{hidden\_dimension} features while the sparse retriever relies on a much smaller activated feature dimension with non-negative learned token importance. On the other hand, after the finetuning phase, \ours achieves performance comparable to its dense counterpart (56.0 versus 56.8). This result suggests the synergy of the two training phases in our training curriculum: the pre-finetuning phase adapts the pretrained causal language model for bidirectional information and sparsity pattern, laying the foundation for the subsequent finetuning; and the high-signal finetuning phase enhances the retriever's ability to identify fine-grained, more nuanced query-document relevance patterns, which is critical for retrieval performance.

\section{Conclusion and Future Work}
In this paper, we presented \ours, a family of learned sparse retrieval models that demonstrate the effectiveness of scaling learned sparse retrieval to LLM backbones. By combining a bidirectional adaptation of \llama with a targeted two-phase training curriculum, we have shown that sparse retrieval can achieve performance parity with dense models while maintaining superior efficiency in index search. \oursEightB currently stands as the most performant sparse retriever on the MTEB Retrieval leaderboard as of February 5th, 2026, offering a scalable solution for high-precision retrieval on commodity hardware. We view these results as evidence that efficiency and scale need not be opposing forces in neural retrieval, and that LLM-based architectures can produce sparse, index-friendly representations without sacrificing effectiveness, pointing toward a unified design space that bridges neural and classical IR. More broadly, we envision a new generation of retrieval systems in which large language models serve not only as generators or re-rankers, but as efficient first-stage retrievers—enabling scalable, high-quality access to information across domains, modalities, and deployment environments.

Future work will explore several promising directions. In particular, we aim to extend \ours to multilingual and multimodal settings~\citep{nguyen2025milcolearnedsparseretrieval}, further optimize the training data mixture to improve performance, and investigate inference-free learned sparse retrieval~\citep{geng2025competitivesearchrelevanceinferencefree,nardini2025effectiveinferencefree,datologyai2025luxicalhighspeedlexicaldensetext}. In parallel, improving end-to-end system efficiency—particularly through faster query encoding and integration with inference optimization techniques—remains important for real-world deployment~\cite{xu-etal-2024-beyond-perplexity}, alongside continued benchmarking against emerging retrieval models and strengthening evaluation practices as the field evolves.

\newcounter{balancebib}
\let\oldbibitem\bibitem
\def\bibitem{\stepcounter{balancebib}\ifnum\value{balancebib}=51 \vfill\eject\fi\oldbibitem}
\bibliographystyle{ACM-Reference-Format}
\bibliography{references}

\end{document}